\documentclass[reprint,prl,superscriptaddress]{revtex4-1}
\usepackage{graphicx}
\usepackage{bm}
\usepackage{natbib}
\def\apjl{Astrophys. J. Lett.}

\def\mnras{Mon.Not.Roy.As.Soc.}

\def\aap{Astron. \& Astrophys.}
\def\ltsima{$\; \buildrel < \over \sim \;$}
\def\simlt{\lower.5ex\hbox{\ltsima}}
\def\gtsima{$\; \buildrel > \over \sim \;$}
\def\simgt{\lower.5ex\hbox{\gtsima}}
\graphicspath{{plots/}}

\def\[{\begin{equation}}
\def\]{\end{equation}}

\renewcommand\section[1]{\emph{#1}.---}
\def\m@th{\mathsurround=0pt }
\def\eqalign#1{\null\,\vcenter{\openup1\jot \m@th
 \ialign{\strut\hfil$\displaystyle{##}$&$\displaystyle{{}##}$\hfil
 \crcr#1\crcr}}\,}

\begin{document}
\title{Clipping the Cosmos: \\ The Bias and Bispectrum of Large Scale Structure}

\author{Fergus Simpson}
\affiliation{SUPA, Institute for Astronomy, University of Edinburgh, Royal Observatory, Blackford Hill, Edinburgh EH9 3HJ}

\author{J. Berian James}
\affiliation{Dark Cosmology Centre, Juliane Maries Vej 30, 2100 Copenhagen \O, Denmark}
\affiliation{Astronomy Department, 601 Campbell Hall, Berkeley CA 94720, USA}

\author{Alan  F. Heavens}
\affiliation{SUPA, Institute for Astronomy, University of Edinburgh, Royal Observatory, Blackford Hill, Edinburgh EH9 3HJ}

\author{Catherine Heymans}
\affiliation{SUPA, Institute for Astronomy, University of Edinburgh, Royal Observatory, Blackford Hill, Edinburgh EH9 3HJ}

\date{\today}
\newcommand{\ud}{\mathrm{d}}
\newcommand{\fpe}{f_\perp}
\newcommand{\fpa}{f_\parallel}
\newcommand{\om}{\Omega_m}
\newcommand{\dmax}{{\delta^{\rm{max}}}}
\newcommand{\Veff}{{V_{\rm{eff}}}}
\newcommand{\eff}{{\rm{eff}}}
\newcommand{\lcdm}{$\Lambda$CDM }
\newcommand{\hmpc}{ \, h \rm{ Mpc}^{-1}}
\newcommand{\hinvmpc}{ \, h^{-1} \rm{ Mpc}}
\newcommand{\tripleint}{\int \! \! \int \! \! \int}
\newcommand{\kfund}{k_f}
\newcommand{\kmax}{k_{\rm{max}}}
\newcommand{\hmpcvol}{ \, h^3 \rm{ Mpc}^{-3}}

\date{\today}

\begin{abstract}
A large fraction of the information collected by cosmological surveys is simply discarded to avoid lengthscales which are difficult to model theoretically. We introduce a new technique which enables the extraction of useful information from the bispectrum of galaxies well beyond the conventional limits of perturbation theory.  Our results strongly suggest that this method increases the range of scales where the relation between the bispectrum and power spectrum in tree-level perturbation theory may be applied, from $\kmax \sim 0.1 \hmpc$  to $\sim 0.7 \hmpc$. This leads to correspondingly large improvements in the determination of galaxy bias.  Since the clipped matter power spectrum closely follows the linear power spectrum, there is the potential to use this technique to probe the growth rate of linear perturbations and confront theories of modified gravity with observation.
\end{abstract}

\maketitle

\section{Introduction}
At the largest observable scales, density fluctuations in the Universe have sufficiently low amplitude to allow a simple approximation to model their evolution, in the form of linear perturbation theory. But at progressively smaller scales this formalism breaks down, and more complex prescriptions are required to predict the clustering patterns. While in recent years there has been steady progress improving our understanding of the nonlinear matter power spectrum (e.g. \cite{1996MNRAS.280L..19P, 2003MNRAS.341.1311S, RPTCrocceScoccimarro}), such developments are unlikely to keep pace with the  precision demanded by the ambitious galaxy surveys of the future. If we are to take full advantage of their capabilities, these surveys  will require knowledge of the matter power spectrum to sub-percent precision at lengthscales well into the nonlinear regime.

The galaxy bispectrum is able to offer insight into key topics in cosmology.  Along with a theoretical model of its evolution, a measurement of the bispectrum enables properties of the galaxy bias to be determined \cite{2002MNRAS.335..432V}. It also potentially holds a signature of inflation, in the form of primordial non-Gaussianity. However, like the power spectrum, the bispectrum is known to be strongly influenced by nonlinearities. To obtain robust parameter constraints, it is usual to discard information above a limiting $\kmax$, an obviously undesirable state of affairs. At low redshifts we are restricted to working on extremely large scales,  typically $\kmax \lesssim 0.1 \hmpc$ \cite{2010Sefusatti}, where cosmic variance severely hampers our precision. This value of $\kmax$ may be improved with more detailed modelling, going beyond tree-level perturbation theory to one-loop or Renormalised Perturbation Theory \cite{RPTCrocceScoccimarro, 2010Sefusatti}; see also \cite{2008PhRvD..78b3523S}. These methods are able to extend the $k$-range of applicability by a moderate amount over tree-level, at the cost of considerable complexity.  An alternative to modelling the full matter power spectrum is to manipulate the real space density field such that the effects of nonlinear growth are suppressed, an approach that has previously been demonstrated to successfully enhance the baryon acoustic oscillations (BAO)  \cite{EisSeoRecon}. There is, therefore, motivation to study such transformations as a general tool for cosmological analysis.

The simplest class of reconstruction methods are local, monotonic mappings from the initial to final density fields. The Gaussianisation process proposed by Weinberg \cite{1992MNRAS.254..315W} enforces Gaussianity in the field's one-point distribution, and this has been shown to recover the shape of the linear power spectrum \cite{Neyrinck2010}. Some other filters, including $\log (1+\delta)$ \cite{Neyrinck2009}, perform comparably well. What such transformations have in common is the penalisation of the highest density regions of the field, a feature we shall exploit further. Our goal is to extend the range of the relatively simple tree-level perturbation theory by manipulating the field in real space. This work begins with the ansatz that the nonlinear behaviour of the density field  is localised not only at high wavenumbers in Fourier space, but also within high density regions in real space. 

\section{Peak Clipping}
Extreme peaks of the density field correspond to the regions furthest from the quasi-linear evolution we wish to study, yet it is these peaks that dominate the contribution to higher-order statistics.  Therefore, a promising approach for extracting the primordial or tree-level gravitational bispectrum is to `clip' the density field, reducing  the most extreme peaks within a sample. We maintain a continuous density field by reducing the value of extreme density peaks down to a threshold value, such that we generate the truncated field $\delta_t (x)$ 

\[ \label{eq:trunc}
\eqalign{
\delta_t(x) &= \delta(x) , \, \, \, \, (\delta(x) < \dmax) \cr
\delta_t(x) &= \dmax, \, \, \, (\delta(x) \ge \dmax)
}
\]

\noindent where the density contrast is defined by $\delta(x) \equiv \rho(x)/\bar{\rho}(x) - 1$. We then re-evaluate $\delta_t(x)$ using the truncated mean density to ensure $\langle \delta_t(x) \rangle = 0$. This prescription allows us to generate results which are rather insensitive to the choice of threshold $\dmax$, provided that it is low enough to affect at least $\sim 0.1\%$ of the volume. 

Galaxy redshift surveyors are well accustomed to dealing with effects that arise from the process of astronomical observation. For measurements of statistics in Fourier space, these are conventionally eliminated by deconvolution of a window function---yet here the effective mask cannot be treated in this manner, as it is strongly correlated with the underlying density field. Crucially, however, the mask covers such a small fraction of the survey volume that we do not need to correct for it in this work. 


\begin{figure*}
\includegraphics[width=180mm]{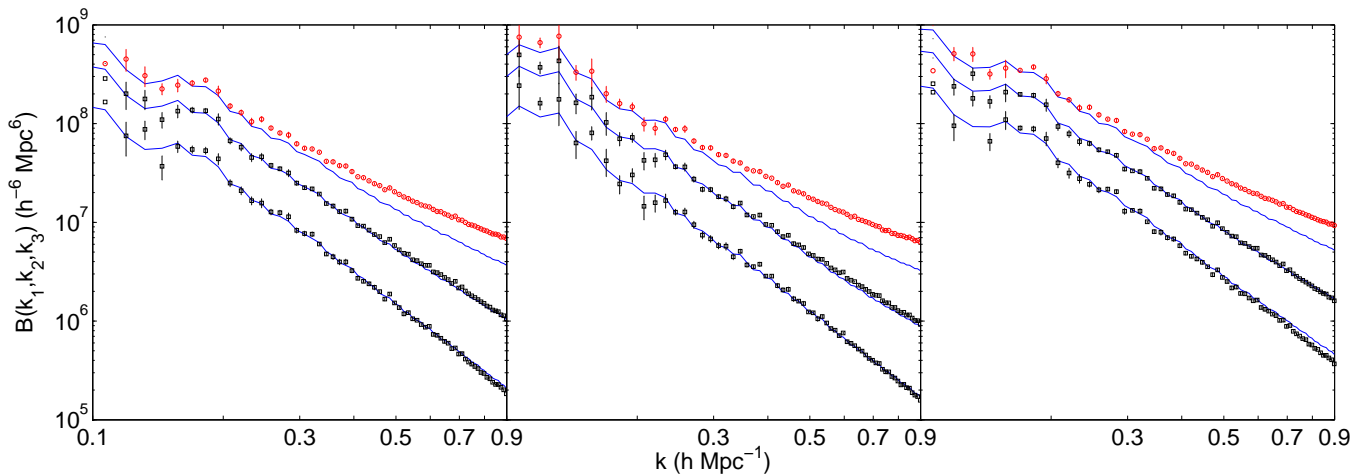}
\caption{ \label{fig:bispec} \textit{Left:} The left and middle panels show the dark matter bispectrum (\ref{eq:bispec}) for colinear triangles with wave vectors in the ratio 3:2:1 and 2:1:1 (left and middle). The solid lines are estimates of the bispectrum using the matter power spectrum, as given by (\ref{eq:bibias}). The upper set  of points and line relates to the untouched density field, while the lower points and lines corresponds to a field which has first been clipped in accordance with (\ref{eq:trunc}). The truncation limits are chosen such that $0.1\%$ (middle) and $1\%$ (lower) of the volume is clipped.   \textit{Right:} The same as the left-hand panel, but here we study the $\log_{10}(M_\ast/M_\odot h) \ge 9$ galaxy sample.}
\end{figure*}

\section{Bispectrum} 
We begin with the assumption that on large scales the number density of galaxies is well modelled in terms of the local linear and quadratic bias parameters $b_1$ and $b_2$, defined such that $\delta_g(x) = b_1 \delta(x) + \frac{b_2}{2} \delta^2(x)$,  where $\delta$ and $\delta_g$ denote the dark matter and galaxy fields respectively.

In a given sample of halos or galaxies, we can attempt to recover the bias properties by comparing the observed bispectrum $B_{g}$ with the tree-level theoretical expectation $\hat{B}_{g}$. These two functions are (see e.g. \cite{Fry1984, 1997MatarreseVHeavens, 1998Scoccimarro, 2010AdAst2010E..73L})

\[ \label{eq:bispec}
 \langle \delta_g(\vec{k}_1) \delta_g(\vec{k}_2) \delta_g(\vec{k}_3) \rangle \, = (2\pi)^3 B_g(k_1, k_2, k_3) \delta(\vec{k}_1+\vec{k}_2+\vec{k}_3)
\] 

\[ \label{eq:bibias}
\hat{B}_{g}(k_1, k_2, k_3) = \frac{1}{b_1} \sum_{i \neq j} \left[ 2 J(k_i, k_j) + \frac{b_2}{b_1} \right] P_g(k_i) P_g(k_j) \, ,
\]

\noindent where $P(k) \equiv \langle |\delta(k)|^2 \rangle $, and the galaxy power spectrum is $P_g(k) = b^2_1 P(k)$. $J(k_1, k_2) $ is given by

\[
J (k_1,k_2) = 1 - D(\om)  + \frac{\cos \theta}{2} \left(\frac{k_1}{k_2} + \frac{k_2}{k_1} \right) + D(\om)  \cos^2 \theta \, ,
\]
where $\theta$ denotes the angle  between $\vec{k}_1$ and $\vec{k}_2$.
This has a weak dependence on cosmology.  For a flat \lcdm  Universe, $D$ may be modelled by $D(\om) \simeq \frac{1}{2} - \frac{3}{14} \om^{-1/143}$ \cite{1995A&A...296..575B, BernardeauScoccimarro2002}. 


\section{Numerical Methods}
We use data from the Millennium-I simulation, utilising both the dark matter density field and an associated catalogue of $\sim 7$ million galaxies \cite{2005Natur.435..629S, Guo2011}. The redshift zero snapshot presents a strongly evolved density field (with $\sigma_8 = 0.9$), providing a suitably challenging test for this method; galaxies are selected by the mass cuts $\log_{10}(M_\ast/M_\odot h) \ge 9$ or $ \ge 10$. These cuts yield number densities of $5.8 \times 10^{-2} \hmpcvol$ and $1.3 \times 10^{-2} \hmpcvol$ respectively.

A galaxy number density field is constructed in the same manner as the dark matter density field,  at the same resolution ($256^3$; box size $= 500 \hinvmpc$) and by the Nearest Grid Point (NGP) algorithm. This field is filtered in real space in accordance with (\ref{eq:trunc}), with $\delta^\mathrm{max}$ chosen such that $0.1\%$ of the volume is clipped. We correct for the window function imposed by the NGP method (see e.g. \cite{Jing})

\[ \label{eq:NGP}
\delta(k) =  \delta_0(k) \prod_{i=x,y,z} \left[ \frac{k_i}{2 \sin \left( \frac{k_i}{k_{\rm{nyq}}} \frac{\pi}{2} \right) } \right]^p \, ,
\]

\noindent where $k_{\rm{nyq}}$ is the Nyquist frequency, and $p=1$ for NGP.

The fundamental frequency of the box $( \kfund \sim 0.013 \hmpc )$ dictates the bin width for estimating our statistics. We perform a brute-force evaluation of all triangle configurations up to $k_{max}= 72 \kfund \simeq 0.9 \hmpc$, storing the cumulative values as a function of the vector amplitudes rounded to the nearest multiple of $\kfund$. 

\[
C(k_a,k_b,k_c) = \sum_{\vec{k}_a}{\sum_{\vec{k}_b}{ \delta(\vec{k}_a) \delta(\vec{k}_b) \delta(\vec{k}_c)}}  \, ,
\]

\noindent where $\vec{k}_c = - \left( \vec{k}_a + \vec{k}_b \right)$. This raw cumulative count is rapidly distilled to any desired configuration

\[ \label{eq:bispec2}
B_g(k_1, k_2, k_3) = \frac{1}{n_t}\sum_{a,b,c \in \rm{tol}} C(k_a,k_b,k_c) \, .
\]

\noindent where we define $k_1 \geq k_2 \geq k_3$, and $n_t$ is the total number of triangles counted. This is evaluated alongside $P(k_1)$, $P(k_2)$, $P(k_3)$ and the mean configuration $\bar{J}(k_1, k_2)$. We select a tolerance value such that the vector magnitudes were an appropriate proportion, matching the desired values of $k_1/k_2$ and $k_1/k_3$ to within $10\%$.
We assume diagonal covariance of the bispectrum estimates, determined by jackknife resampling from eight subsamples. This is a potentially important issue which we will explore in future work, as it could result in underestimation of the error bars, but the evidence from the bias recovery in Figure \ref{fig:contours} is that it isn't a significant effect in this study

In order to determine whether this technique has successfully recovered the properties of the linearised field, we need to estimate the true values of $b_1$ and $b_2$.  For the dark matter, this is simply given by  $b_1=1$,  $b_2=0$. For the galaxy field, we fix $b_1$ from the ratio of large-scale power spectra, and determine $b_2$ by least-squares fitting to fields smoothed with a cutoff $\kmax = 0.2 \hmpc$.

For a galaxy field $b_1$ and $b_2$ are not invariant under the clipping process. Galaxy bias is sensitive to the selection criteria of the sample, and so we find that for truncation fractions around $\sim 1\%$ the underlying bias values change by up to $\sim 10\%$.  Note that the change of bias is not necessarily important:  a useful by-product of this procedure is to determine the {\em matter} power spectrum, which can be obtained from the clipped galaxy power spectrum and the clipped bias.  An advantage is that the resulting clipped matter power spectrum is very close to the linear power spectrum.

\section{Results} 
Figure  \ref{fig:bispec} displays the impact of clipping on the bispectrum of the  dark matter density field (left and middle panels) and the galaxy number density field (right panel). In each panel, the upper set of data points represent the unclipped bispectrum, while the middle and lower sets of data points show the bispectrum after $0.1\%$ and $1\%$ of the field has been clipped. The solid lines denote the tree-level prediction for the bispectrum. The effect of the clipping is to reduce the extreme nonlinearities, bringing the tree-level prediction into much closer agreement with the data and permitting much higher wavenumbers to be incorporated.  The left and right panels have wave vectors in the ratio 3:2:1, and the middle panel 2:1:1. Beyond $k > 0.7 \hmpc$ the correction given by (\ref{eq:NGP}) becomes large and the results in this regime may not be reliable.

Clipping  $0.1\%$ of the volume corresponds to a threshold of $\delta_{max} \simeq 60$ and $70$ for the dark matter and galaxy fields, corresponding to mass fractions of  $\sim14\%$ (dark matter) and $\sim15\%$ (galaxies).  As the fraction of the volume subject to clipping increases, so does the shot noise, as the number density of galaxies and the amplitude of the filtered power spectrum are both suppressed. Note that coarser grid spacings would require larger fractions of the volume to be clipped, in order to ensure a sufficient number of extreme peaks are affected.

\begin{figure*} 
\includegraphics[width=180mm]{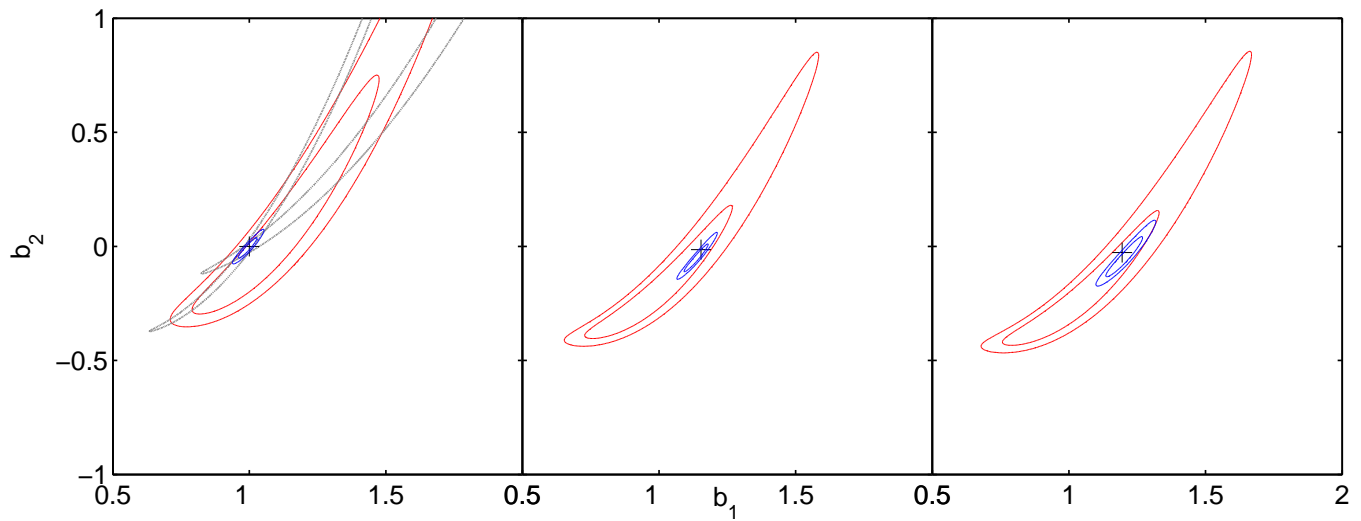} 
\caption{\label{fig:contours} Constraints on the linear and quadratic bias parameters, as derived from combining two configurations of the bispectrum, before (red) and after (blue) clipping $0.1\%$ of the volume.  \textit{Left:} This is for the $z=0$ dark matter density field, so the correct values are $b_1=1$ and $b_2=0$, as denoted by a black cross. For the flat configuration, the clipping allows us to approximately triple the Fourier cut $\kmax$ from $0.17 \hmpc$ to $0.5 \hmpc$. The dotted contours outline the individual contributions for the case of the filtered field. \textit{Middle and Right:} Same as the left panel, except here we are using a mock galaxy catalogue with mass cuts $\log_{10}(M_\ast/M_\odot h) \ge 9$ and $\ge 10$ respectively.} 
\end{figure*}

Any given configuration $B(k_1,k_2,k_3)$ exhibits degeneracy between the parameters $b_1$ and $b_2$. Figure \ref{fig:contours} illustrates how a combination of two configurations (3:2:1 and 3:2:2) helps to lift this degeneracy.  We compute the likelihood of $b_1$ and $b_2$ from equation (\ref{eq:bibias}), assuming that the errors in the power spectrum are negligible in comparison with the bispectrum.  The red and blue contours are derived from unclipped and clipped $(0.1\%)$ density fields respectively, and represent the one- and two- $\sigma$ confidence contours. The clipping allows us to reach a higher $\kmax$, substantially reducing the errors on the bias parameters. The left panels correspond to the dark matter field, while the middle and right panels are for galaxy fields with mass cuts $\log_{10}(M_\ast/M_\odot h) \ge 9$ and $\ge 10$ respectively. 

For the original unclipped field we take  $\kmax = 0.17 \hmpc$. If we attempt to enlarge this value, the confidence contours rapidly diverge from the true value. When working with the filtered field, we are able to utilise the  bispectrum up to $0.5 \hmpc$ for the scalene flattened (3:2:1) shape. The isosceles (3:2:2) is slightly less responsive to the clipping process, and for that configuration we apply a cut at $\kmax = 0.35 \hmpc$. We note that higher truncation fractions appear to allow even higher values of  $\kmax$ for the co-linear configurations. Triangle configurations closer to the equilateral shape suffer from a decrement at larger truncation fractions. We speculate this may be due to the co-linear Fourier triangles picking out pancake-like structures, which are lower density and thus less susceptible to clipping than the filaments, to which equilateral configurations are more sensitive. It is expected that this effect could be calibrated using N-body simulations. 

We have also repeated these tests with a higher redshift $(z=0.687)$ dark matter field, effectively probing the dependence on the power spectrum. We find similar results for these cases, that clipping just $0.1\%$ of the field significantly extends the reach of (\ref{eq:bibias}). We investigated lognormal mapping and Gaussianisation, which can be effective at power spectrum level \cite{1992MNRAS.254..315W,Neyrinck2010,Neyrinck2009}; however at bispectrum level we did not find good agreement with (\ref{eq:bibias}).

\section{Conclusions}
Determining the form of the dark matter power spectrum remains a principal goal of modern cosmology. We have demonstrated how the simple prescription of clipping high density peaks allows a much larger volume of $k$-space to be incorporated into such predictions. This leads to a substantially more accurate determination of the bias, and consequently tighter constraints on cosmological parameters.  For the highly evolved field of the Millennium simulation ($\sigma_8 = 0.9$ at redshift zero), we find that the maximum wavenumber $\kmax$ may typically be extended by this technique from $\sim 0.1 \hmpc$ to beyond $\sim 0.5 \hmpc$. The full impact  of this extension upon signal to noise becomes apparent upon noting that the number of triangles contributing to the bispectrum computations scales in proportion to $\kmax^6$. 

In this letter we have chosen to focus on the galaxy bias parameters, but it should be clear that this technique is extensible to quite generic measurements in cosmology. At the heightened precision with which the matter bispectrum can now be studied, it is natural to consider whether improved constraints on $f_{NL}$ are attainable.  Although any signature of primordial non-Gaussianity is expected to be most prominent at the largest scales, the abundance of modes at high $k$ may compensate for the smaller signal. The clipping procedure may also prove valuable in improving cosmological constraints on the neutrino mass \cite{2008PhRvL.100s1301S, 2009PhRvD..79b3520I, 2009PhRvD..80h3528S, 2010MNRAS.405..168L}, and in modified gravity models where the linear growth rate differs from General Relativity.

There are a number of ways in which the technique could be developed further.  A closer-to-optimal approach is to evaluate the relationship between the preferred truncation threshold and particular wavenumber values. Indeed the ultimate limit on $\kmax$ is at present uncertain, since larger truncation fractions extend the range of validity beyond $0.7 \hmpc$. The compensation of the clipping window, and moving to redshift space, where there will undoubtedly be challenges due to `Fingers of God', will be topics of forthcoming work. Exploitation of the bispectrum to the level of precision that will soon be observationally accessible will likely require the incorporation of scale-dependence into the bias model. Considering these future directions, it seems reasonable to conclude that density filtering schemes may establish themselves as an important tool for cosmological analysis.

We thank Peder Norberg, Sylvain de la Torre, Benjamin Joachimi and Thomas Kitching for helpful discussions. FS and AFH acknowledge the Dark Cosmology Centre's visitor programme. FS and CH acknowledge support from the European Research Council under the EC FP7 grant number 240185. JBJ is supported by the Danish National Research Foundation's Sophie \& Tycho Brahe Programme in Astrophysics. 

\bibliography{/Volumes/katrine.roe.ac.uk/Routines/dis}
\end{document}